\magnification=1200
\input iopppt.modifie
\input epsf
\def\received#1{\insertspace 
     \parindent=\secindent\ifppt\textfonts\else\smallfonts\fi 
     \hang{#1}\rm } 
\def\appendix{\goodbreak\beforesecspace 
     \noindent\textfonts{\bf Appendix}\secspace} 
\def\figure#1{\global\advance\figno by 1\gdef\labeltype{\figlabel}%
   {\parindent=\secindent\smallfonts\hang 
    {\bf Figure \ifappendix\applett\fi\the\figno.} \rm #1\par}} 
\headline={\ifodd\pageno{\ifnum\pageno=\firstpage\titlehead
   \else\rrhead\fi}\else\lrhead\fi}

\def\rrhead{\textfonts\hskip\secindent\it 
    \shorttitle\hfill\rm L\folio} 
\def\lrhead{\textfonts\hbox to\secindent{\rm L\folio\hss}%
    \it\aunames\hss} 
\footline={\ifnum\pageno=\firstpage
\hfil\textfonts\rm L\folio\fi}   
\def\titlehead{\smallfonts J. Phys. A: Math. Gen. {\bf 24} (1991)
L1119--L1125  \hfil} 

\firstpage=1119
\pageno=1119

\jnlstyle
\jl{1}
\overfullrule=0pt

\letter{Statistics of nested spiral
self--avoiding loops:\hfil\break  
exact results on the square and
triangular lattices}[Letter to the Editor]

\author{L Turban}[Letter to the Editor]
 
\address{Laboratoire de Physique du Solide\footnote{\dag}{Unit\'e de
Recherche associ\'ee au CNRS no 155}, Universit\'e de Nancy I, BP 239
\hfil\break F--54506 Vand\oe uvre l\`es Nancy Cedex, France}

\received{Received 3 April 1991}

\abs
The statistics of nested spiral self--avoiding loops, which is closely
related to the partition of integers into decreasing parts, has been
studied on  the square and triangular lattices. The number of
configurations with $N$ steps is $c_N\simeq (\sqrt{2}/24)N^{-3/2}
\exp(\pi\sqrt{\frac{2}{3}}N^{1/2})$ and their  average size $X_N\simeq
(1/2\pi)\sqrt{\frac{3}{2}}N^{1/2}\ln N$ to leading order on the
square lattice while the corresponding values for the triangular
lattice are $c_N\simeq(3^{3/4}/16)N^{-5/4}
\exp((\pi/\sqrt{3})N^{1/2})$ and $X_N\simeq1/(\pi\sqrt{3})N^{1/2}\ln
N$.     
\endabs

\vglue1cm


\submitted
\date

Some years ago, the number of $N$-step spiral
self--avoiding loops have been  calculated for the square
and triangular lattices (Manna 1985, Lin et al 1986).
These works followed the introduction of the spiral
self--avoiding walk (Privman 1983) for which a lot of exact
results were obtained by a succession of authors (Bl\"ote
and Hilhorst 1984, Whittington 1984, Gutmann and Wormald
1984, Joyce 1984, Guttmann and Hirschhorn 1984, Lin 1985,
Joyce and Brak 1985, Lin and Liu 1986). While the number
of spiral self--avoiding loops grows with $N$ like a
nonuniversal, i.e. lattice--dependent power, the number of
spiral self--avoiding walks also behaves in an unusual way
$$
C_N\simeq AN^{-\theta} \exp {(\lambda N^{1/2})}
\eqno(1)
$$
where both $\theta$ and $\lambda$ are
lattice--dependent quantities. It follows that the
asymptotic entropy per step decays  as $N^{-1/2}$ instead
of giving a constant like in the ordinary or directed
self--avoiding walks. Following the work of Privman, a
close connection between this problem and the theory of
partitions of integers (Andrews 1976) was noticed (Derrida
and Nadal 1984,  Redner and de Arcangelis 1984, Klein et
al 1984).

In this letter we present some exact results concerning
the statistics of  nested spiral loops which are self-- and
mutually avoiding and piled up around  a site chosen as
origin. We study such spirals on the square lattice where
on a loop only $90^\circ$ turns in the same direction are
allowed so that the loops are rectangular--shaped (figure 1)
and on the triangular lattice where the restriction to
$120^\circ$ turns leads to triangles among which one only
keeps those pointing up (figure 2).
{\par\begingroup\parindent=0pt\medskip
\epsfxsize=9truecm
\topinsert
\centerline{\epsfbox{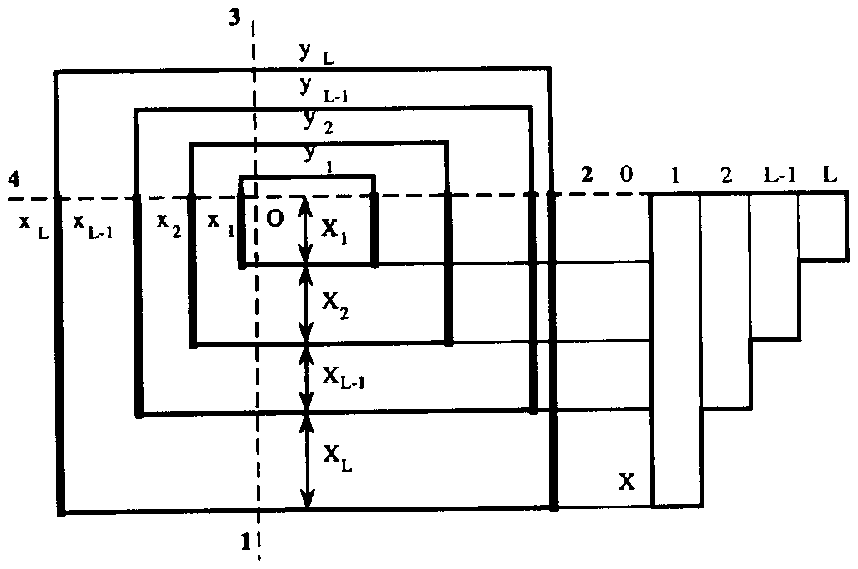}}
\smallskip
\figure{Nested spiral self--avoiding loops on the square lattice:
with $90^\circ$ turns in the same direction, rectangular--shaped loops
are obtained. Each step is assigned a weight $z$ and the size is
$X\!=\!\sum_{k=1}^LX_k$. The nested--loop configuration corresponds to
four independent partitions of integers into decreasing
parts numbered $1$ to $4$ and each partition is duplicated (heavy
lines).}  
\endinsert 
\endgroup
\par}
{\par\begingroup\parindent=0pt\medskip
\epsfxsize=9truecm
\midinsert
\centerline{\epsfbox{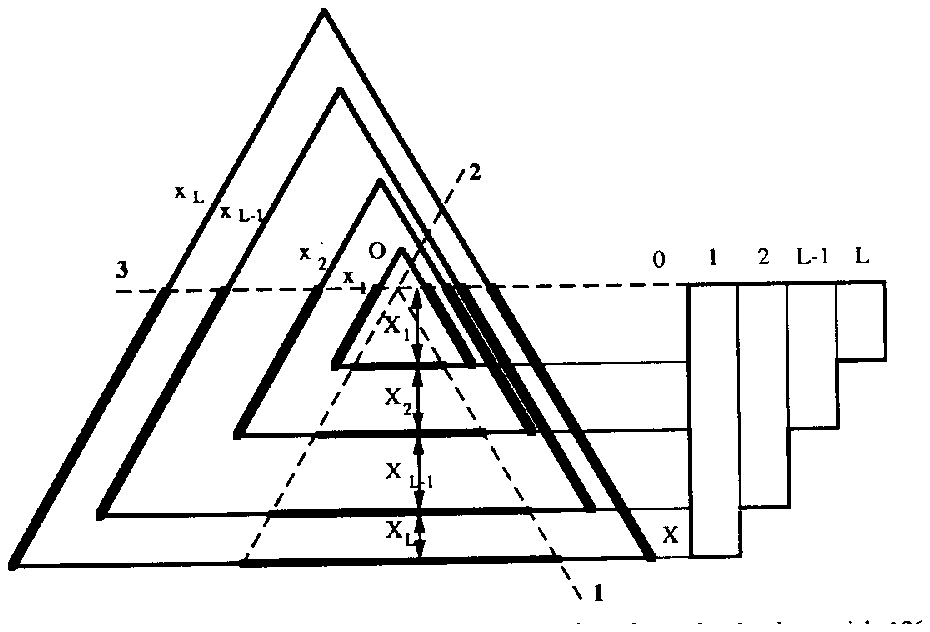}}
\smallskip
\figure{Nested spiral self--avoiding loops on the triangular lattice:
with $120^\circ$ turns in the same direction, triangular--shaped loops
are obtained. Each step is assigned a weight $z$ and the size is
$X\!=\!\sum_{k=1}^LX_k$. The nested--loop configuration corresponds to
three independent partitions of integers into decreasing parts
numbered $1$ to $3$ and each partition is triplicated (heavy
lines).}   
\endinsert 
\endgroup
\par}

Let us introduce the generating function
$$
G_L(z,\omega )=\sum_{N,X}c_N(L,X)z^Ne^{\omega X}
\eqno(2)
$$ 
for the number of configurations with $N$ steps,
$L$ loops and size $X\!=\!\sum_{k=1}^LX_k$ where X is the
distance from the origin to the $L$th loop.  On the
square lattice, with the notations of figure 1,
one may write
$$
\fl G_L^{sq}(z,\omega )=\sum_{x_1=2}^\infty\sum_{y_1=2}^\infty (y_1-1)
\sum_{X_1=1}^{x_1-1}\sum_{x_2=x_1+2}^\infty
\sum_{y_2=y_1+2}^\infty (y_2-y_1-1) \sum_{X_2=1}^{x_2-x_1-1}\cdots\cr
\fl\qquad\cdots \sum_{x_L=x_{L-1}+2}^\infty\sum_{y_L=y_{L-1}+2}^\infty
(y_L-y_{L-1}-1) \sum_{X_L=1}^{x_L-x_{L-1}-1}
z^{2\sum_{k=1}^L(x_k+y_k)}e^{\omega
\sum_{k=1}^LX_k}
\eqno(3)
$$
Introducing
$$
m_k=x_k-x_{k-1}\qquad x_0=0\eqno(4a)\cr
n_k=y_k-y_{k-1}\qquad y_0=0
\eqno(4b)
$$
one obtains
$$
\sum_{k=1}^L(x_k+y_k)=\sum_{k=1}^L(L-k+1)(m_k+n_k)
\eqno(5)
$$ 
and the generating function may be factorized
as
$$
\fl G_L^{sq}(z,\omega)=\prod_{k=1}^L
\left[\sum_{m_k=2}^{\infty} 
z^{2(L-k+1)m_k}\left({e^\omega -e^{m_k\omega }\over
1-e^\omega}\right)  
\sum_{n_k=2}^{\infty}(n_k-1)z^{2(L-k+1)n_k}\right]
\eqno(6)
$$
or
$$
G_L^{sq}(z,\omega )=\prod_{k=1}^L
{e^\omega z^{8k}\over(1-z^{2k})^3
(1-z^{2k}e^\omega)}
\eqno(7)
$$
In the same way, on the triangular lattice with the
notations of figure 2, one  has

$$
\fl G_L^{tr}(z,\omega )=\sum_{x_1=3}^{\infty}
\sum_{X_1=1}^{x_1-2}(x_1-X_1-1) \sum_{x_2=x_1+3}^{\infty}
\sum_{X_2=1}^{x_2-x_1-2}(x_2-x_1-X_2-1)\cdots\cr
\cdots \sum_{x_L=x_{L-1}+3}^{\infty}
\sum_{X_L=1}^{x_L-x_{L-1}-2}
(x_L-x_{L-1}-X_L-1)z^{3\sum_{k=1}^Lx_k}e^{\omega\sum_{k=1}^LX_k}
\eqno(8)
$$ 
so that using (4a)
$$
\fl G_L^{tr}(z,\omega)=\prod_{k=1}^L
\left[\sum_{m_k=3}^{\infty}
{(m_k-2)e^\omega-(m_k-1)e^{2\omega}+
e^{\omega m_k}\over (1-e^\omega )^2}
z^{3(L-k+1)m_k}\right]
\eqno(9)
$$
and finally
$$
G_L^{tr}(z,\omega )=\prod_{k=1}^L
{e^\omega z^{9k}\over (1-z^{3k})^2(1-e^\omega
z^{3k})}
\eqno(10)
$$

These results may be obtained more directly by noticing
the connection with the  number of partitions of integers
into $L$ decreasing parts which is illustrated on
figure 1 for the square lattice and figure 2 for the
triangular lattice. Nested loops configurations
are in one--to--one correspondence with four 
independent partitions in the first case and
three independent partitions in  the second
case. The generating functions for the
nested--loop problem are directly obtained as
powers of the generating function for the
partition of integers into $L$ decreasing parts
(Andrews 1976)
$$\eqalign{
g_L(t)=&(t+t^2+t^3+\cdots)(t^2+t^4+t^6+\cdots)\cdots 
(t^L+t^{2L}+t^{3L}+\cdots )\cr
=&\prod_{k=1}^L{t^k\over 1-t^k}\cr}
\eqno(11)
$$
multiplied by a modified generating function
for which a partition of size $X$ is, as in
equations~(3) and~(8), weighted by $e^{\omega X}$ so 
that 
$$\eqalign{
g_L(t,\omega)=&(t e^\omega+t^2 e^{2\omega}+\cdots) (t^2 e^\omega 
+t^4 e^{2\omega}+\cdots)\cdots(t^L e^\omega+t^{2L} e^{2\omega
}+\cdots)\cr =&\prod_{k=1}^L{e^\omega t^k\over1-e^\omega t^k}\cr}
\eqno(12)
$$ 
One may verify that the mapping requires $t\!=\! z^2$
for loops on the square lattice (figure~1) and
$t\!=\! z^3$ on the triangular lattice (figure~2); it
follows that
$$
G_L(t,\omega)=\left[g_L(t)\right]^{p-1}
g_L(t,\omega )|_{t=z^q}
\eqno(13)
$$ 
with $p\!=\!4$, $q\!=\!2$, on the square lattice and $p\!=\!3$, $q\!=\!3$,
on the triangular lattice in agreement with equations~(7) and~(10).

Let us first consider the generating function $G(t)$ for
the number of  configurations $c_N\!=\!\sum_{L,X}c_N(L,X)$
with $N$ steps and any number of  loops which is given by
$$
G(t)=\sum_{L=1}^{\infty}G_L(t,\omega )|_{\omega =0}=\sum_{L=1}^{\infty}
\left[g_L(t)\right]^p|_{t=z^q} 
\eqno(14)
$$
The behaviour of $c_N$ for large $N$ values is governed by
the behaviour of $g_L(t)$ in the vicinity of its singularity
at $t\!=\!1$. With $t\!=\!1\!-\!\eta$ and $\eta\!\rightarrow\!0^+$, the
main  contribution to the sum in equation (14) comes from values of $L$ near
$L_0\!=\!\ln 2/\eta$ for which $g_L(\eta )$ is maximum in $L$
(Bl\"ote and Hilhorst 1984, des Cloizeaux and Jannink 1987) 
and an expansion of $\ln g_L(\eta )$ near $L_0$ (see the appendix)
leads to
$$
\ln g_L(\eta )\simeq {\pi^2\over 12\eta }-{1\over 2}
\ln \left({2\pi \over \eta }\right) -{1\over \eta} (L\eta
-\ln 2)^2
\eqno(15)
$$
so that the sum over $L$ in $G(t)$ may be transformed into
a Gaussian integral  and one obtains
$$
G(t)=\sum_{L=1}^{\infty} \exp [p\ln g_L(t)]\simeq
(2p)^{-p/2}\left({p\eta \over \pi}\right)^{p-1\over
2}\exp \left({p\pi^2\over 12\eta}\right)
\eqno(16)
$$
for the leading contribution. With $n\!=\! N/q$, the generating
function may be  written as
$$
G(t)=\sum_{n=0}^{\infty}c_{N=qn}t^n
\eqno(17)
$$
so that the number of configurations is given by the
Cauchy formula
$$
c_{N=qn}={1\over 2\pi i}
\oint_{(C)}dt{G(t)\over t^{n+1}} 
\eqno(18)
$$
where $(C)$ is a circle of radius $r\!<\!1$ 
centered at the origin. The integral may be evaluated
using the saddle--point method by deforming the contour.
Through a Gaussian integration near the saddle--point at
$t\!=\!1\!-\!\pi(p/12n)^{1/2}$ one obtains
$$
c_N\simeq {(pq)^{p/4}(q/p)^{1/2}\over 2^{p+1}3^{p/4}}
N^{-{p+2\over 4}} 
\exp \left(\pi \sqrt{p\over 3q}N^{1/2}\right)
\eqno(19)
$$
to leading order. With the appropriate $p$
and $q$ values one obtains
$$
c_N^{sq}\simeq {\sqrt 2\over 24}N^{-3/2}
\exp \left(\pi \sqrt{2\over 3}N^{1/2}\right)
\eqno(20a)\cr
c_N^{tr}\simeq {3^{3/4}\over16}N^{-5/4} 
\exp \left({\pi \over \sqrt{3}}N^{1/2}\right)
\eqno(20b)
$$
For both lattices, the exponential term is the same as 
for the outward spiral self--avoiding walk (Bl\"ote and
Hilhorst 1984, Joyce 1984, Guttmann and  Wormald 1984,
Joyce and Brak 1985, Lin and Liu 1986) whereas the power
of $N$ in the prefactor is different.

According to equation (2), the mean size for $L$-loop
configurations is given by
$$
X_L(t)={\partial \ln G_L(t,\omega )\over 
\partial \omega}\bigg|_{\omega=0} =\sum_{k=1}^L{1\over 1-t^k} 
\eqno(21)
$$
and may be rewritten as
$$
X_L(t)=\sum_{k=0}^{\infty} {t^k(1-t^{Lk})\over
1-t^k}
\eqno(22)
$$
where the first term in the sum should be understood as
the limit of the ratio when $k\!\rightarrow\!0$. When $L$ is
unrestricted, the  mean size becomes
$$
X(t)={\sum_{L=1}^{\infty} G_L(t)X_L(t)\over G(t)}
=\sum_{k=0}^{\infty} {t^k \over 1-t^k}
{\sum_{L=1}^{\infty} G_L(t)(1-t^{Lk})\over G(t)}
\eqno(23)
$$
Changing the sum over $L$ into an integral,
the value $L_0\!=\!\ln{2}/\eta$ corresponding to the
maximum in $L$ of $G_L(t)$ is selected and
$$
X(t)\simeq \sum_{k=0}^{\infty}
{t^k(1-2^{-k})\over 1-t^k}
\eqno(24)
$$
Putting apart the first term and rearranging the sum, one
obtains
$$
X(\eta )\simeq {\ln 2\over \eta }+\sum_{k=1}^{\infty}
\left({1\over 1-t^k}- {1\over 1-t^k/2}\right)
\eqno(25)
$$
The sum may be evaluated using the Euler--Maclaurin formula
and
$$
X(\eta )\simeq {1\over \eta}\ln ({1\over \eta })
\eqno(26)
$$
to leading order. On the other hand the number
of steps reads
$$
N(\eta )={\partial \ln G(z^q)\over \partial
\ln{z}}\bigg|_{z^q=1-\eta} 
\simeq {pq\pi^2\over 12\eta^2}
\eqno(27)
$$
so that
$$
X_N\simeq {1\over 2\pi }\sqrt{12\over pq}
N^{1/2}\ln{N}
\eqno(28)
$$
and
$$
X_N^{sq}\simeq {1\over 2\pi }\sqrt{3\over 2}
N^{1/2}\ln{N}
\eqno(29a)\cr
X_N^{tr}\simeq {1\over \pi \sqrt{3}}
N^{1/2}\ln{N}
\eqno(29b)
$$
One recovers the characteristic behaviour of the spiral
self--avoiding walk (Bl\"ote and Hilhorst 1984, Liu and Lin
1985) with an exponent $\nu\!=\!1/2$ and a logarithmic
correction.

\appendix

Using equation (11) with $t\!=\!1\!-\!\eta$ in the limit 
$\eta\!\rightarrow\!0^+$ one has
$$
\fl\ln {g_L(t)}=-\sum_{k=1}^L\ln{(t^{-k}-1)}\simeq -\sum_{k=1}^L
\left[\ln{(e^{k\eta}-1)}-\ln{k\eta}\right]-\sum_{k=1}^L\ln{k\eta}
\eqno(A1)
$$
The first sum may be replaced by an integral using
$$
\sum_{k=1}^Lf(k)\simeq {1\over \eta}\int_{\eta\over 2}^{(L+{1\over
2})\eta}f(u) du +{\eta \over 24}\left[f'({\eta \over 2})-f'\left((L+{1\over
2})\eta \right)\right]Ê
\eqno(A2)
$$
so that
$$
\fl\ln{g_L(\eta)}=-{1\over \eta}\int_0^{(L+{1\over 2})\eta}
\left[\ln{(e^u-1)}-\ln u\right] du-L\ln \eta-\ln {(L!)}+O(\eta)
\eqno(A3)
$$
Integrating the second term and using Stirling formula, one gets to leading
order
$$ 
\ln{g_L(\eta)}\simeq -{1\over \eta}\int_0^{(L+{1\over 2})\eta}
\ln{(e^u-1)} du-{1\over 2}\ln{\left({2\pi \over \eta}\right)}
\eqno(A4)
$$
Expanding the remaining integral, considered as a function of its upper limit,
near $u_0\!=\!L_0\eta\!=\!\ln2$ with
$$
\int_0^{\ln 2}\ln{(e^u-1)}du=-{\pi^2\over 12}
\eqno(A5)
$$
one finally obtains
$$
\ln {g_L(\eta )}\simeq {\pi^2\over 12\eta }-{1\over 2}
\ln \left({2\pi \over \eta }\right) -{1\over \eta} (L\eta
-\ln 2)^2
\eqno(A6)
$$
\vfill \eject

\references

\refbk{Andrews G E  1976}{The Theory of
Partitions}{(London: Addison--Wesley)}
 
\refjl{Bl\"ote H W J and Hilhorst H J 1984}{\JPA}{17}{L111} 

\refbk{des Cloizeaux J and Jannink G 1987}{Les Polym\`eres en
Solution}{(Les Ulis: Les Editions de Physique) p~65}
 
\refjl{Derrida B  and Nadal J  P  1984}{J. Phys.
Lett.}{45}{L701} 

\refjl{Guttmann A J and Hirschhorn M 1984}{\JPA}{17}{3613}  

\refjl{Guttmann A J and Wormald N C 1984}{\JPA}{17}{L271} 

\refjl{Joyce G S 1984}{\JPA}{17}{L463} 

\refjl{Joyce G S and Brak R 1985}{\JPA}{18}{L293}	

\refjl{Klein D J, Hite G E,Schmalz T G and Seitz
W A  1984}{\JPA}{17}{L209}

\refjl{Lin K V 1985}{\JPA}{18}{L145}

\refjl{Lin K V, Chiu S H, Ma S K and Kao C H 1986}{\JPA}{19}{3093}

\refjl{Lin K V and Liu K C 1986}{\JPA}{19}{585}

\refjl{Liu K C and Lin K V 1985}{\JPA}{18}{L647}

\refjl{Manna S S 1985}{\JPA}{18}{L71}

\refjl{Privman V 1983}{\JPA}{16}{L571}

\refjl{Redner S and de Arcangelis L 1984}{\JPA}{17}{L203}

\refjl{Whittington S G 1984}{\JPA}{17}{L117}

\vfill\eject
\bye